\documentclass[
aps,
floatfix,
letterpaper,
prx,
singlecolumn,
reprint,
superscriptaddress]{revtex4-2}

\usepackage[sectionbib]{bibunits}
\defaultbibliographystyle{apsrev4-2} 
\defaultbibliography{Correlated_sensing} 

\usepackage[colorlinks=true, linkcolor=blue]{hyperref}

\usepackage{lipsum}
\usepackage{graphicx} 
\usepackage{amsfonts}
\usepackage{amsmath}
\usepackage{braket}
\usepackage{gensymb}
\usepackage{hyperref}
\usepackage{dsfont}
\usepackage[capitalize]{cleveref}

\usepackage[utf8]{inputenc}

\newcommand{\Sx}{S_1}
\newcommand{\Sy}{S_2}
\newcommand{\sx}{s_1}
\newcommand{\sy}{s_2}
\newcommand{\sxi}{s_{1,i}}
\newcommand{\syi}{s_{2,i}}

\newcommand{\sigmaSx}{\sigma_{S_{1}}}
\newcommand{\sigmaSy}{\sigma_{S_{2}}}

\newcommand{\phix}{\phi_1}
\newcommand{\phiy}{\phi_2}
\newcommand{\nx}{n_1}
\newcommand{\ny}{n_2}

\newcommand{\rxy}{r}
\newcommand{\rideal}{r_\text{ideal}}
\newcommand{\rPC}{r_\text{pc}}
\newcommand{\rSS}{r_\text{th}}

\newcommand{\seq}{\mathord{=}}

\newcommand{\Xms}{m_{s_1}}
\newcommand{\Yms}{m_{s_2}}

\newcommand{\Phix}{\Phi_1}
\newcommand{\Phiy}{\Phi_2}

\newcommand{\phiGx}{\phi_{C_1}}
\newcommand{\phiGy}{\phi_{C_2}}
\newcommand{\phiGxy}{\phi_{C_{1,2}}}
\newcommand{\phiG}{\phi_C}
\newcommand{\phiLx}{\phi_{L1}}
\newcommand{\phiLy}{\phi_{L2}}
\newcommand{\phiL}{\phi_{L}}

\newcommand{\sigmaG}{\sigma_{\Phi_C}}
\newcommand{\e}{\textrm{e}}

\DeclareMathOperator{\Pois}{Pois}

\date{\today}

\begin{document}
\begin{bibunit}[apsrev4-2]

\begin{abstract}
    Nitrogen vacancy (NV) centers in diamond are atom-scale defects with long spin coherence times that can be used to sense magnetic fields with high sensitivity and spatial resolution. Typically, the magnetic field projection at a single point is measured by averaging many sequential measurements with a single NV center, or the magnetic field distribution is reconstructed by taking a spatial average over an ensemble of many NV centers. In averaging over many single-NV center experiments, both techniques discard information. Here we propose and implement a new sensing modality, whereby two or more NV centers are measured simultaneously, and we extract temporal and spatial correlations in their signals that would otherwise be inaccessible. We analytically derive the measurable two-point correlator in the presence of environmental noise, quantum projection noise, and readout noise. We show that optimizing the readout noise is critical for measuring correlations, and we experimentally demonstrate measurements of correlated applied noise using spin-to-charge readout of two NV centers. We also implement a spectral reconstruction protocol for disentangling local and nonlocal noise sources, and demonstrate that independent control of two NV centers can be used to measure the temporal structure of correlations. Our covariance magnetometry scheme has numerous applications in studying spatiotemporal structure factors and dynamics, and opens a new frontier in nanoscale sensing.
\end{abstract}

\title{Nanoscale covariance magnetometry with diamond quantum sensors}
\author{Jared Rovny} 
\affiliation{Princeton University, Department of Electrical and Computer Engineering, Princeton, NJ 08544, USA}
\author{Zhiyang Yuan} 
\affiliation{Princeton University, Department of Electrical and Computer Engineering, Princeton, NJ 08544, USA}
\author{Mattias Fitzpatrick} 
\thanks{Present address: Thayer School of Engineering, Dartmouth College, Hanover, NH 03755, USA}
\affiliation{Princeton University, Department of Electrical and Computer Engineering, Princeton, NJ 08544, USA}
\author{Ahmed I.\ Abdalla} 
\thanks{Present address: Department of Electrical Engineering, Stanford University, Stanford, CA 94305, USA}
\affiliation{Princeton University, Department of Electrical and Computer Engineering, Princeton, NJ 08544, USA}
\author{Laura Futamura} 
\thanks{Present address: Department of Physics, Stanford University, Stanford, CA 94305, USA}
\affiliation{Princeton University, Department of Electrical and Computer Engineering, Princeton, NJ 08544, USA}
\author{Carter Fox} 
\affiliation{University of Wisconsin-Madison, Department of Physics, Madison, WI 53706, USA}
\author{Matthew Carl Cambria}
\affiliation{University of Wisconsin-Madison, Department of Physics, Madison, WI 53706, USA}
\author{Shimon Kolkowitz}
\affiliation{University of Wisconsin-Madison, Department of Physics, Madison, WI 53706, USA}
\author{Nathalie P.\ de Leon}
\thanks{Corresponding author. Email: npdeleon@princeton.edu}
\affiliation{Princeton University, Department of Electrical and Computer Engineering, Princeton, NJ 08544, USA}

\maketitle

\section{Introduction}

Correlated phenomena play a central role in condensed matter physics, and have been studied in many contexts including 
phase transitions \cite{Bernien2017,Zhang2017},
many-body interactions and entanglement \cite{Cheneau2012,Shankar2017,Altman2004,Deng2005,Baez2020}, and
magnetic ordering \cite{Simon2011,Mazurenko2017}, 
as well as in the context of fluctuating electromagnetic fields, where two-point correlators are central to characterizing field statistics \cite{Lifshitz1980,Joulain2005,Premakumar2017,Agarwal2017}. Recent efforts towards improving quantum devices have also explored correlated noise in SQUIDS \cite{Sendelbach2009,Yoshihara2010,Gustavsson2011} and qubits \cite{Szankowski2016,Viola2017,Krzywda2019,vonLupke2020,Wilen2021,Tennant2022}.
Nitrogen vacancy (NV) centers in diamond are a promising sensing platform for detecting correlations, as they are robust, noninvasive, and capable of measuring weak signals with nanoscale resolution \cite{Casola2018}. These advantages have made them a useful tool for studying many condensed matter systems including magnetic systems like 2D van der Waals materials \cite{Thiel2019,Sun2021}, magnons \cite{LeeWong2020}, and skyrmions \cite{Dovzhenko2018,Yu2018,Jenkins2019}; and transport phenomena like Johnson noise \cite{Kolkowitz2015}, hydrodynamic flow \cite{Vool2021,Ku2020,Jenkins2022}, and electron-phonon interactions in graphene \cite{Anderson2019}. These applications are powerful but have so far been limited to signals that are averaged over space or time --- more information is potentially available by studying spatial and temporal correlations in the system. Significant advances in nanoscale spectroscopy have already been made by studying correlations from a single NV center at different points in time \cite{Laraoui2013,Boss2017,Pfender2019}; measuring correlated dynamics between two different NV centers would provide simultaneous information at length scales ranging from the diffraction limit to the full field of view ($\sim$0.1--100 micron length scales). Furthermore, measuring two NV centers allows for measurements of correlations at two different sensing times limited only by the experimental clock cycle ($\sim$1 ns resolution). Measurements of spatiotemporal correlations at these length and time scales would provide useful information about the dynamics of the target system, including the electron mean free path, signatures of hydrodynamic flow \cite{Levitov2016}, or the microscopic nature of local NV center noise sources like surface spins \cite{Romach2015,Sangtawesin2019,Dwyer2021}. 

In this paper we develop a new technique to measure classical correlations between two noninteracting NV centers, which gives access to nonlocal information that would normally be discarded with single NV center measurements. Measuring such two-point correlators with NV centers is challenging because conventional optical spin readout provides very little information per shot. Here we derive the sensitivity requirements for detecting correlations, and experimentally implement a covariance magnetometry protocol using spin-to-charge readout of two spatially separated NV centers to achieve low readout noise. We demonstrate correlation measurements of random-phase classical magnetic fields measured at two points separated in space and time, and implement a spectral decomposition method for extracting and distinguishing between correlated and uncorrelated spectral components.

\begin{figure*}[ht]
	\centering
	\includegraphics[width=4.75in]{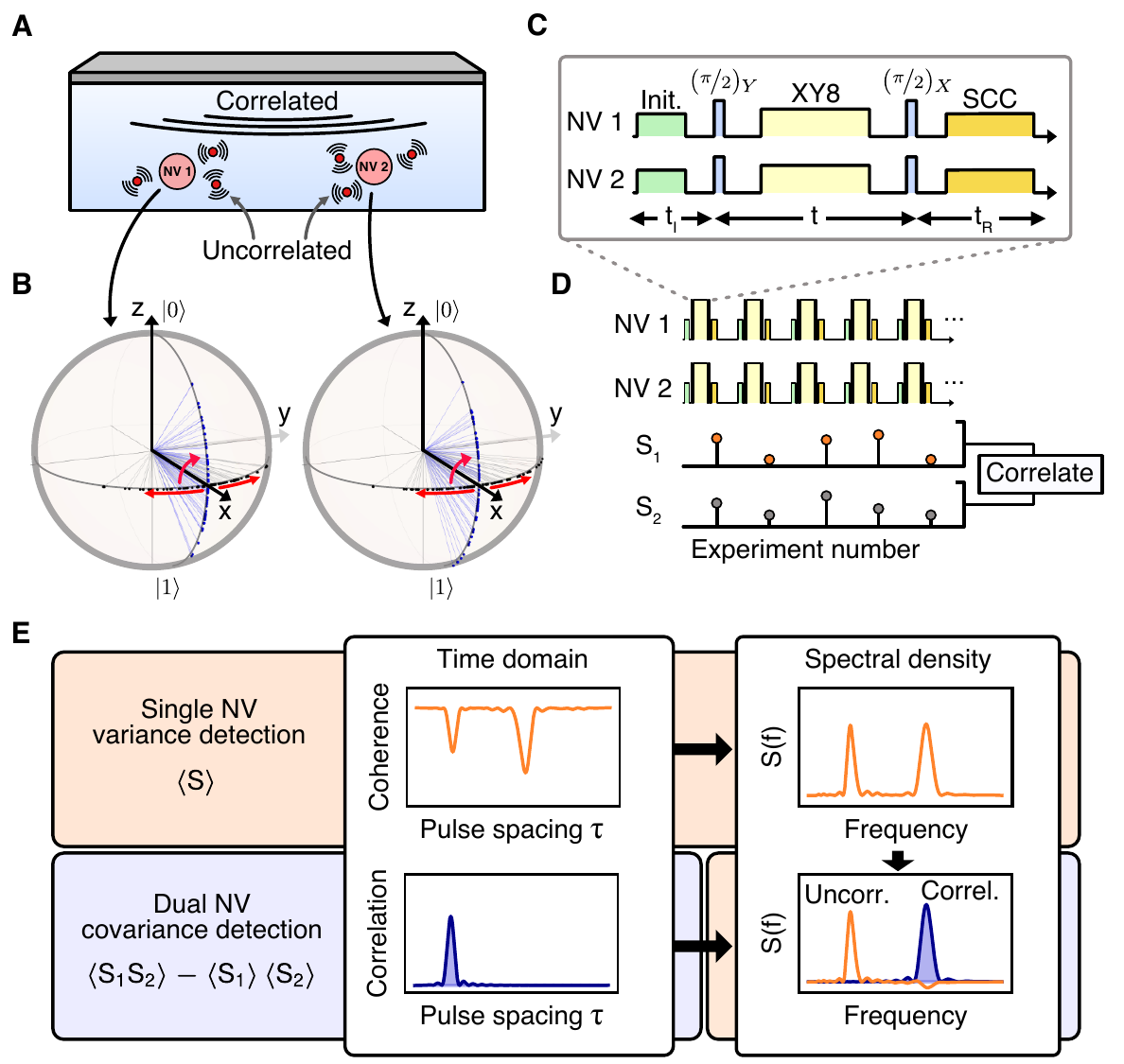}
	\caption{Covariance noise sensing. (A) Diagram of a diamond with two near-surface NV centers experiencing uncorrelated local fields and a correlated common field. (B) Bloch sphere representations of each qubit state during sensing, with the states prepared along $x$ followed by a phase accumulation which will be different in each experiment, resulting in a distribution of phases. At the end of each experiment a final $\pi/2$ pulse maps these phases to populations. (C-D) Pulse sequence diagrams showing the sensing (XY8) and measurement (SCC) sequence for each NV center. The measurement is repeated many times, retaining the photon counts from each measurement without signal averaging; we instead measure the correlation between the resulting lists $S_i$. (E) Using conventional detection of single NV centers (top row), the coherence decay gives access to the noise spectral density $S(f)$ but provides no spatial information. Covariance magnetometry measuring two NV centers (bottom row) provides information about which spectral features are correlated and which are uncorrelated.}
	\label{fig:overview}
\end{figure*}

We consider two NV centers that do not directly interact with each other but experience a shared classical magnetic field, whose amplitude is correlated at the locations of the two NV centers (\cref{fig:overview}A). Each NV center also sees a unique local magnetic field that is uncorrelated between the two locations. These fields are detected using a Ramsey-type experiment addressing the $m_s=0$ and $m_s=+1$ (or $-1$) spin sublevels of the NV center (referred to as states 0 and 1 respectively), as illustrated in \cref{fig:overview}B-D. Upon many repeated measurements, we accumulate a list of signals $\Sx=\{\sxi\}$ and $\Sy=\{\syi\}$, where $i=1...N$ indexes the $N$ total experiments.

Though similar to a typical Ramsey-type variance detection sequence \cite{Degen2017}, we emphasize two significant modifications for covariance detection. First, despite detecting zero-mean noise, we choose a final pulse that is 90 degrees out of phase with the initial pulse, such that for high-frequency noise detection the final spin state is equally likely to be 0 or 1 (\cref{fig:overview}B,C), maximizing our sensitivity to correlations. This is not done in conventional noise detection using variance magnetometry, since straightforward signal averaging would then produce the same result $\braket{m_{s_i}}=0.5$ always. Second, we do not compute the average value of this signal, but rather compute the shot-to-shot cross-correlation between the raw signals $\Sx$ and $\Sy$ (\cref{fig:overview}D).

\begin{figure*}[ht]
	\centering
	\includegraphics[width=7in]{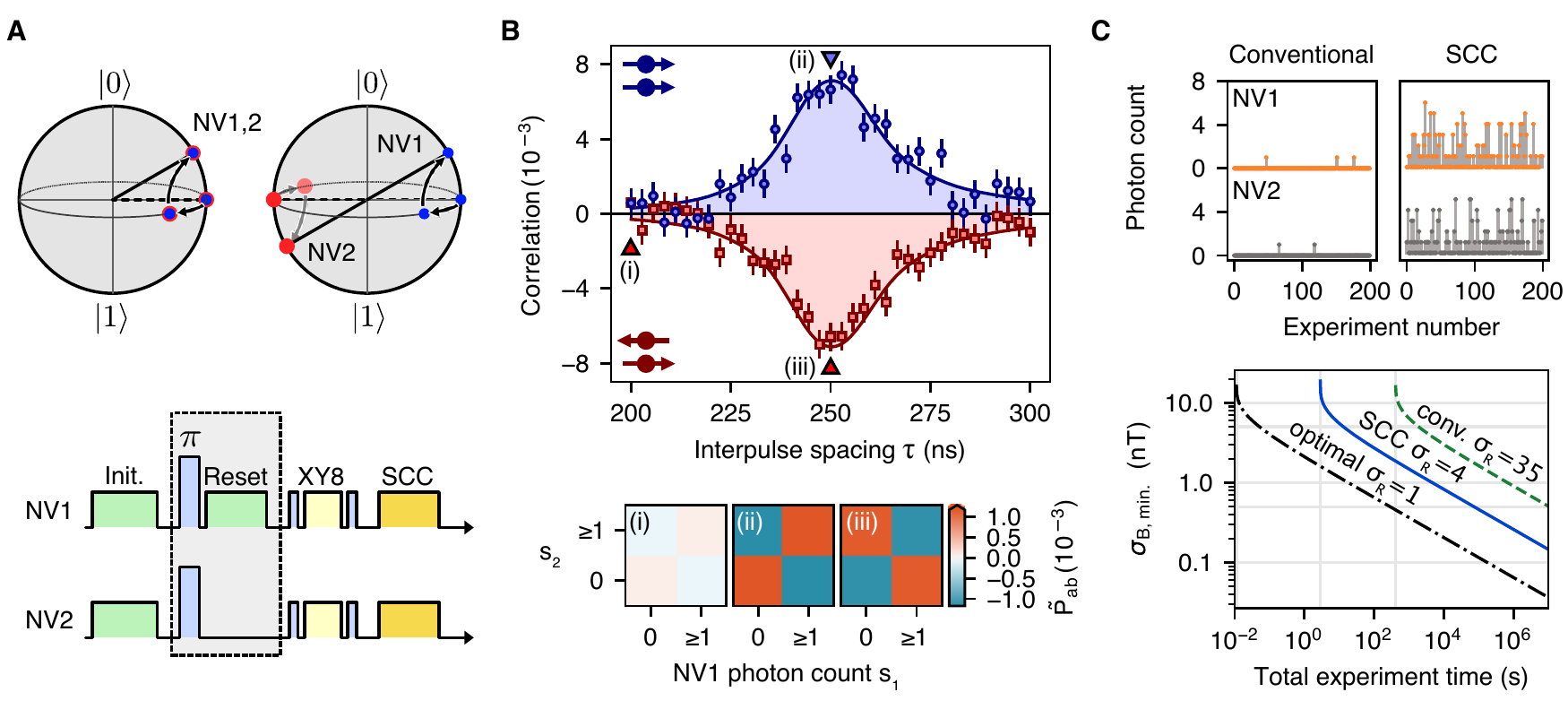}
	\caption{Detecting correlations and anticorrelations. (A) Pulse sequence and final Bloch sphere mapping for correlation (top left) or anticorrelation (top right) measurements using global microwave control. For anticorrelations, an extra $\pi$ pulse and spatially selective NV polarization optical pulse (``reset'') are added during initialization (bottom, gray box). (B) Correlation detected from a 2 MHz AC signal whose phase is randomized with 1 MHz bandwidth Gaussian noise. The measured correlations are positive when the NV centers are initialized parallel to one another (blue circles) and negative when they are initialized antiparallel (red squares). Lines indicate the predicted correlation shape \cite{Supp}. Raw photon count statistics (bottom) taken from the marked data points in the top panel show no correlation (i), positive correlation (ii), or negative correlation (iii), where the color indicates the joint detection probability $\tilde{P}_{ab} \equiv P(s_1\seq a,s_2\seq b)-P(s_1\seq a)P(s_2\seq b)$. (C) Comparison of shot-to-shot photon counts during averaging for conventional readout (top left) and spin-to-charge conversion readout (top right). (bottom) Minimum magnetic field amplitude to detect correlations with $\text{SNR}=1$ for Gaussian noise. Here we have assumed $T_2=100\,$ $\mu$s and the phase integration time $t=T_2/2=50\,$ $\mu$s, as well as a readout time of $300\,$ns for conventional readout and $1\,$ms for SCC and optimal readout. Initialization time was ignored.
	}
	\label{fig:ACcorrelations}
\end{figure*}

Whereas conventional variance measurements provide spectral densities with no spatial information (\cref{fig:overview}E, top row), the addition of covariance information allows us to identify which spectral components are common between two NV centers and which are unique to each (\cref{fig:overview}E, bottom row). Throughout this work, we focus on the measured Pearson correlation $\rxy = \text{Cov}(\Sx,\Sy)/(\sigma_1 \sigma_2)$, where Cov is the covariance and $\sigma_{1,2}$ is the standard deviation of $S_{1,2}$.

\section{Detecting correlations}

To demonstrate our protocol, we use an external radiofrequency (RF) coil or stripline to apply a global, random phase AC signal to two shallow NV centers approximately 10 nm from the diamond surface. Here the two NV centers share the same magnetic resonance frequency, so all microwave pulses address both. They are spatially resolved, allowing for separate excitation and readout using two independent optical paths \cite{Supp}. To boost the sensitivity of our readout, we use a simultaneous spin-to-charge conversion (SCC) protocol \cite{Shields2015, Barry2020} on each NV center separately. We use an XY8 sensing protocol for each NV center to maximize sensitivity to the applied AC signal \cite{Gullion1990}  (\cref{fig:ACcorrelations}A). As expected, we observe correlations that are maximized when the interpulse spacing matches the frequency of the global signal (\cref{fig:ACcorrelations}B, blue circles). The correlations are apparent in the photon count statistics (\cref{fig:ACcorrelations}B, bottom panel ii); when one or more photons are detected from NV1, we observe a higher likelihood of also detecting a photon from NV2. To confirm that we are in fact detecting correlations in the spin state of the NV centers rather than spurious technical correlations \cite{Supp}, we can also initialize the two NV centers on opposite sides of the Bloch sphere prior to applying the XY8 sequence (\cref{fig:ACcorrelations}A). The phase accumulation step then results in a final state that is anticorrelated between the two NV centers (\cref{fig:ACcorrelations}B, red squares).

\begin{figure*}[ht]
	\centering
	\includegraphics[width=4.6in]{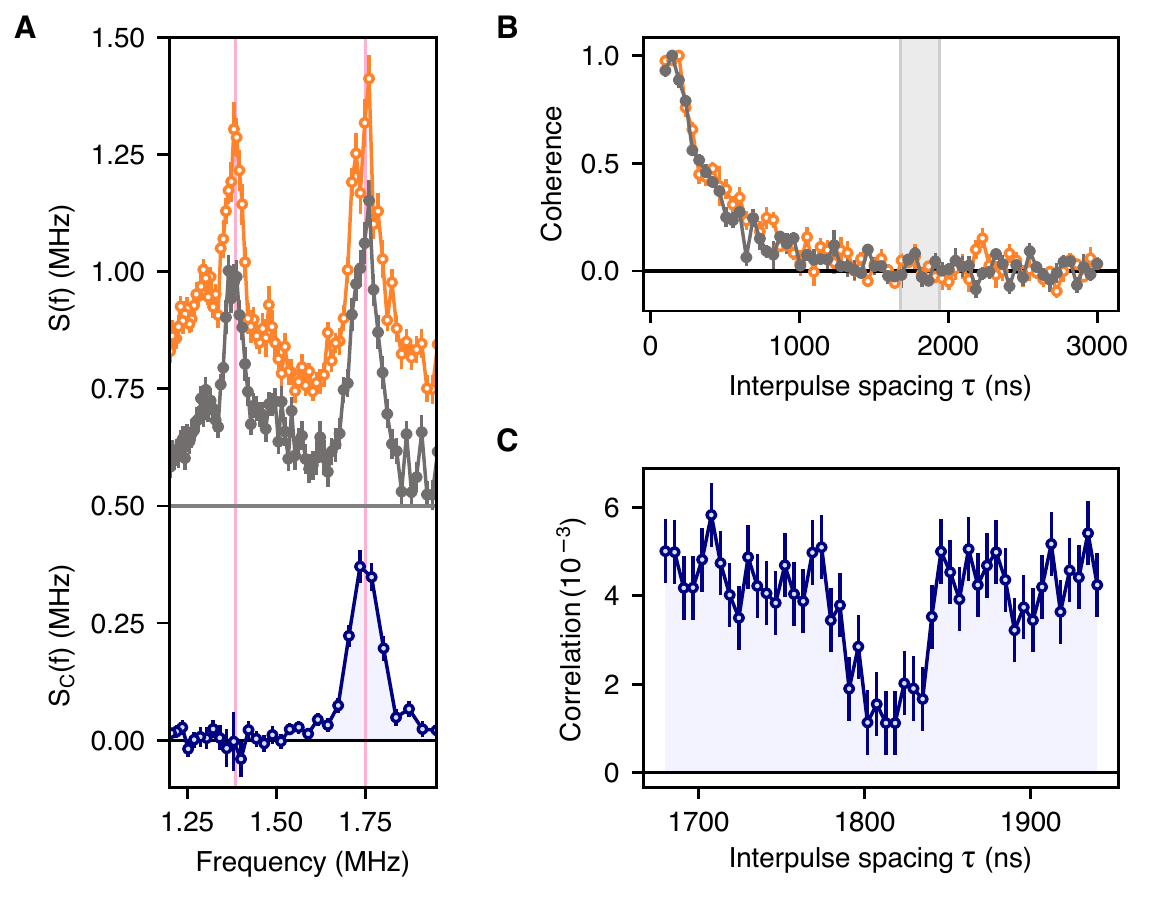}
	\caption{Disentangling correlated and uncorrelated signals. (A) Single-NV noise spectra derived from conventional XY8 variance magnetometry (top) of two NV centers (orange open markers and gray filled markers, arbitrarily offset). Each NV center detects signals at two common frequencies, but it is impossible to directly determine whether the sources are local or nonlocal. Spectral decomposition (bottom) using covariance magnetometry  (\cref{eq:reconstructedSpectrum}) reveals that the higher frequency peak is caused by a shared noise source. Here, the shared noise feature is engineered using an applied global 1.75 MHz AC signal, while the local feature is caused by the $^{15}$N nuclear spin intrinsic to each NV center. (B) In a broadband correlated noise environment, the two NV centers rapidly decohere (orange open markers and gray filled markers). (C) Covariance magnetometry for evolution times indicated by the gray rectangle in (B) reveals a dip in the Pearson correlation around $\tau=1800$ ns arising from the uncorrelated $^{15}$N nuclear spins intrinsic to each NV center. The broadband noise is correlated, allowing for the observation of spectral features from local signals even at evolution times beyond the coherence time of both NV centers.}
	\label{fig:spectraldecomposition}
\end{figure*}

The sensitivity of a covariance measurement differs from that of a traditional magnetometry measurement because it requires simultaneous signals from two NV centers. Assuming that the detected phases are statistically even, as for a noisy or random-phase signal, we find \cite{Supp} the Pearson correlation
\begin{align}
    r = 
    \frac{\e^{-[\tilde{\chi}_1(t_1)+\tilde{\chi}_2(t_2)]}}{\sigma_{R_1}\sigma_{R_2}}  \braket{\sin[\phiGx(t_1)]\sin[\phiGy(t_2)]}, \label{eq:rpc}
\end{align}
where the subscripts $1,2$ denote NV1 and NV2 respectively, the decoherence function $\tilde{\chi}_{1,2}(t)$ describes the `typical' coherence decay of the NV centers due to the local fields \cite{Cywinski2008}, $\phiGxy$ are the phases accumulated by the NV centers due to the correlated field, and the readout noise $\sigma_{R_{1,2}}=\sqrt{1+2(\alpha_0+\alpha_1)/(\alpha_0-\alpha_1)^2}$ characterizes the fidelity of a photon-counting experiment with mean detected photon number $\alpha_0,\alpha_1$ for spin states $0,1$ respectively \cite{Taylor2008}.
For thresholding, the readout noise instead depends on the fidelity of the spin state assignment \cite{Supp}.

\begin{figure*}[ht]
	\centering
	\includegraphics[width=7in]{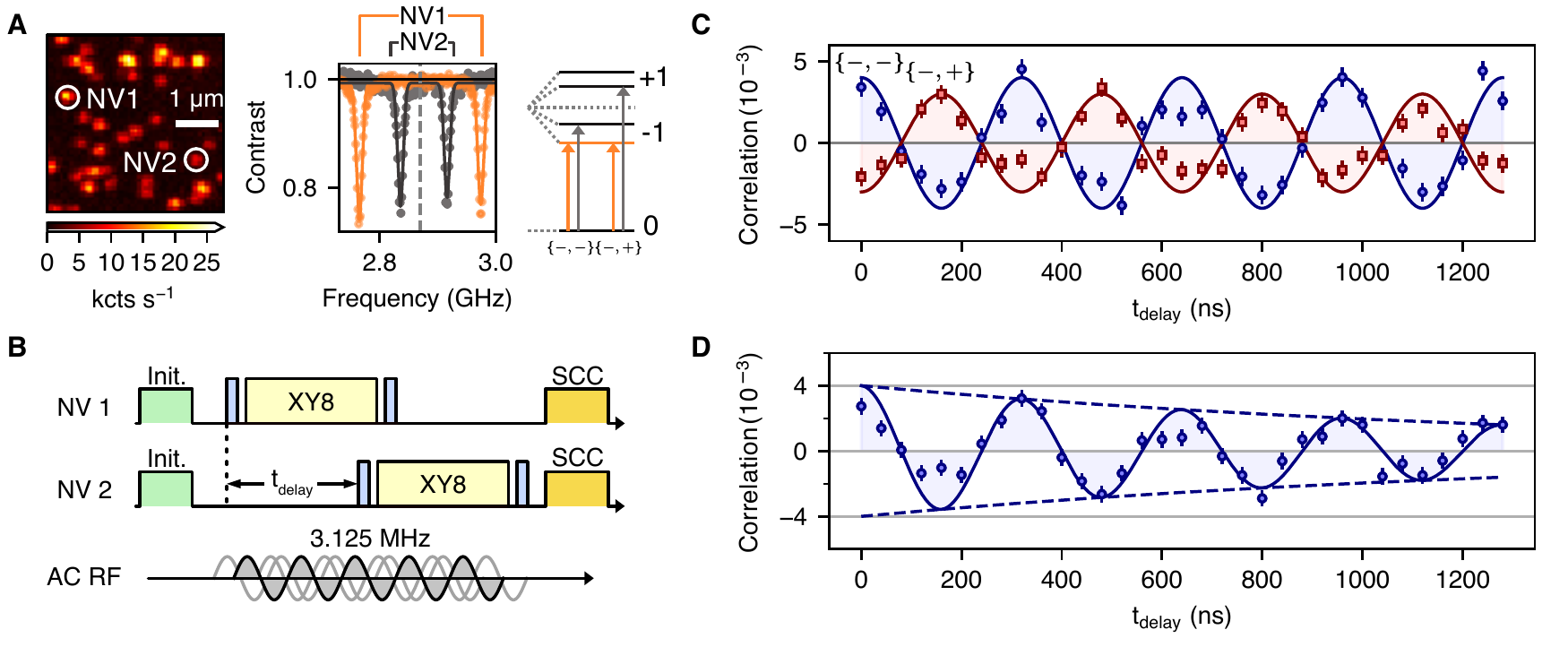}
	\caption{Temporal structure in correlations using independent control. (A) Confocal image showing the two NV centers used for these experiments (left). Optically detected magnetic resonance spectrum (middle) showing optical contrast as a function of microwave drive frequency displays two distinct sets of transitions corresponding to NV1 and NV2, with assignments (right). The NV centers are driven independently on either the $(0,-1)$ transitions for both NVs, labeled $\{-,-\}$, or the $(0,-1)$ and $(0,+1)$ transitions for NV1 and NV2 respectively, labeled $\{-,+\}$. (B) Diagram of the pulse sequence used to probe temporal correlations. After initialization, the start of the XY8 pulse sequence applied to NV2 is delayed by time $t_\text{delay}$ from the start of the pulses on NV1. A $f_0=3.125\,$MHz global AC signal is applied, making the resonant XY8 interpulse spacing $\tau=160\,$ns. (C) Correlations for the case where the NV centers are addressed on the same transitions ($\{-,-\}$, blue circles) oscillate as a function of $t_\text{delay}$ at the AC signal frequency $3.125\,$MHz. The correlations invert (red squares) when the two NV centers are addressed on different transitions ($\{-,+\}$), as they now accumulate opposite phases for the same signal. (D) With added phase noise, the time-domain dephasing of the AC signal is resolvable, despite having a short coherence time (less than $2\,\mu$s) compared to the XY8 sequence time.}
	\label{fig:timeoffset}
\end{figure*}

Note that the detectable correlation depends quadratically on the readout noise, making readout fidelity especially important for detecting correlations; this key fact is implicit in prior calculations of single-NV center two-point correlators derived in the context of repeated weak measurements \cite{Pfender2019}. This may be intuitively understood from \cref{fig:ACcorrelations}C, which shows the raw photon counts for conventional versus SCC readout methods. Using conventional readout, only approximately $0.01$ photons are detected per measurement, such that detecting simultaneous counts from both NV centers is extremely unlikely. Using SCC readout dramatically increases our ability to detect coincident events, and has a greater effect on covariance measurements than on conventional single-NV center measurements. From the independently measured values for each term on the r.h.s.\ of \cref{eq:rpc} \cite{Supp}, we expect the detectable correlation in our experiment to be approximately bounded by $r \approx 0.01$, in good agreement with the maximum correlation $r\approx 0.008$ we detect here (\cref{fig:ACcorrelations}B). The remaining discrepancy is likely due to imperfect charge state initialization and SCC ionization.

Because readout noise plays an amplified role in covariance detection, covariance measurements can become prohibitively long without optimizing sensitivity, for which we require a detailed understanding of the signal to noise ratio (SNR). The sensitivity (minimum noise amplitude $\sigma_{B,\text{min}}$ with $\text{SNR}=1$) of an experiment  detecting Gaussian noise is given by \cite{Supp}
\begin{align}
    \sigma_{B,\text{min}}^2 &= \frac{-\pi\cdot\text{Hz}}{4\gamma_e^2t} \ln \left( 1-\frac{2\sigma_R^2 \e^{2t/T_2}}{\sqrt{T/(t+t_\text{R})}} \right), \label{eq:sensitivitygeneral1}
\end{align}
where $\gamma_e$ is the electron gyromagnetic ratio, $t$ is the phase integration time, $T_2$ is the coherence time, $t_\text{R}$ is the readout time, and $T\approx (t+t_\text{R})N$ is the total experiment time ignoring initialization. This
is shown in \cref{fig:ACcorrelations}C (bottom) for three different readout methods: conventional ($\sigma_R=35$), spin-to-charge conversion ($\sigma_R=4$), and single-shot readout with perfect fidelity ($\sigma_R=1$), which is ultimately limited by quantum projection noise. Achieving $\text{SNR}=1$ for these three scenarios when \ $\sigma_B=1\,$nT requires of order $300\,$ hours, $3\,$ hours, and $10\,$ seconds respectively. 
While detecting correlations is extremely inefficient using conventional readout, enhanced readout protocols like spin-to-charge conversion \cite{Shields2015,Hopper2018,Barry2020,Irber2021,Zhang2021} allow for drastically lower readout noise, making covariance magnetometry possible to implement in practice. 

\section{Disentangling correlated and uncorrelated noise sources}
 
Detecting cross-correlations in pure noise reveals previously hidden information about the spatial structure of the noise, which we now demonstrate using two NV centers sensing both local and nonlocal magnetic fields. We first measure the spectral density $S(f)$ using a conventional variance magnetometry measurement of two different NV centers (\cref{fig:spectraldecomposition}A). These individual spectra reveal that there are two frequencies where signals are seen by both NV centers, but cannot provide simultaneous nonlocal spatial information about that signal. Using covariance magnetometry over the same frequency range (\cref{fig:spectraldecomposition}B) shows only the higher-frequency feature, which clearly reveals that the higher-frequency feature is caused by a noise signal common to each NV center, while the lower-frequency feature is instead caused by local noise sources unique to each NV center.

This ability to distinguish correlated and uncorrelated features enables spatially-resolved spectral decomposition, allowing us to distinguish spectral components that are shared from those that are local. For phases that are Gaussian-distributed or small ($\phi\ll\pi$) we can find \cite{Supp} the correlated noise spectrum $S_C(f)$ if we have access to both the two-NV correlation $r$ as well as each NV center's coherence decay $C_i(t)=\e^{-\chi(t)}$ (note that $C_i(t)$ includes both the correlated and uncorrelated noise sources):
\begin{align}
    S_C(f) = \frac{\pi}{2t}
    \sinh^{-1}\left(\frac{\sigma_R^2 r}{C_1(t) C_2(t)}\right), \label{eq:reconstructedSpectrum}
\end{align}
where $t=n/(2f)$ and $n$ is the total number of applied XY8 pulses. This equation is used to obtain the correlated spectrum from the measured correlation and single-NV center coherence decays, as shown in \cref{fig:spectraldecomposition}A. The local spectrum for each NV center $S_{L_{1,2}}(f)$ may also be found from each individual NV center's total spectrum $S_{L_{1,2}}(f)= S_{1,2}(f)-S_C(f)$.

So far we have analyzed the case where shared and local features are spectrally resolved, but an interesting scenario arises when a shared signal decoheres each NV center at frequencies coincident with local noise sources. In order to probe this case, we apply a global broadband Gaussian noise signal, decohering both NV centers while inducing broadband correlations in their phases (\cref{fig:spectraldecomposition}B-C). Beyond the coherence time of each NV center, conventional variance detection cannot reveal any information (\cref{fig:spectraldecomposition}B, gray region). However, covariance magnetometry (\cref{fig:spectraldecomposition}C) measures the broadband correlation in the random phases of the decohered NV centers --- this correlation will dip if either NV center interacts with a local noise source in its vicinity, as the local signal induces a phase that is unique to that NV center. The covariance magnetometry spectrum therefore reveals a feature that is hidden in the single-NV spectra.

\section{Temporal structure of correlations}

Covariance magnetometry also enables measurements of the temporal structure of the two-point correlator  $\braket{B(r_1,t_1)B(r_2,t_2)}$ separated in time as well as space for short timescales where $t_2-t_1<t+t_R$, which is not possible with single NV center correlation measurements \cite{Laraoui2013,Boss2017,Pfender2019}. To perform this measurement, independent control of each NV center is required. We accomplish this by choosing two NV centers with different orientations at low magnetic fields (\cref{fig:timeoffset}A), such that the $0\rightarrow -1$ transition of the NV center that is aligned with the magnetic field is detuned by $70\,$MHz from that of the misaligned NV center. We then offset the beginning of the XY8 sequence applied to NV2 by time $t_\text{delay}$ (\cref{fig:timeoffset}B), and measure an applied AC field at frequency $f_0=3.125\,$MHz.
As we sweep $t_\text{delay}$, the correlations oscillate at frequency $f_0$ (\cref{fig:timeoffset}C), as expected for a random-phase AC signal \cite{Laraoui2013,Degen2017}. Independent control also allows us to simultaneously address opposite spin transitions for each NV center (\cref{fig:timeoffset}A, right). Since the two NV centers then accumulate opposite phases from the AC field, we observe anti-correlations with the same frequency (\cref{fig:timeoffset}C, red squares). 

Because the two NV centers are manipulated independently, there are no fundamental constraints on the length of $t_\text{delay}$. This allows us to directly measure time-domain structure on the nanosecond time scale at two points in space, despite using $\pi$ pulses with $60\,$ns duration. When we measure the correlations between two NV centers experiencing a shared AC signal with added phase noise (\cref{fig:timeoffset}D), we can directly resolve the temporal structure of the AC signal despite its short coherence time of less than $2\,\mu$s, without making use of spectral deconvolution.

\section{Conclusions and outlook}

Here we have demonstrated simultaneous control and readout of two spatially resolved NV centers, and have shown that enhanced readout enables nanoscale magnetometry of two-point spatiotemporal field correlators that would normally be discarded using conventional NV center magnetometry. This new measurement technique has many potential applications; specifically, measurements of these two-point correlators can reveal the underlying length and time scales of fluctuating electromagnetic fields near surfaces \cite{Lifshitz1980,Joulain2005,Premakumar2017,Agarwal2017}, providing information about nonequilibrium transport dynamics \cite{Dolgirev2022} and condensed matter phenomena like magnetic ordering in low-dimensional systems \cite{Simon2011,Mazurenko2017,Thiel2019}. Future extensions of the current demonstration include using photonic structures to improve photon collection efficiency \cite{Hopper2018,Barry2020}, applying different pulse sequences to each NV center to probe the correlations between signals at different frequencies \cite{Cundiff2013} or phases \cite{Cywinski2008}, and using detector arrays to perform simultaneous readout of many pairs of NV centers. 

\putbib[Correlated_sensing]

\begin{acknowledgments}
We gratefully acknowledge helpful conversations with Alex Burgers, Sarang Gopalakrishnan, and Jeff Thompson. \textbf{Funding:} Developing the covariance sensing protocol and shallow NV center preparation was supported by the NSF under the CAREER program (grant DMR-1752047) as well as the Princeton Catalysis Initiative, and spin-to-charge readout and charge state stabilization was supported by the US Department of Energy, Office of Science, Office of Basic Energy Sciences, under Award No. DE-SC0018978. Work performed at UW-Madison was supported by the U.S. Department of Energy Office of Science National Quantum Information Science Research Centers. J.R. acknowledges the Princeton Quantum Initiative Postdoctoral Fellowship for support. M.F. acknowledges the Intelligence Community Postdoctoral Research Fellowship Program by Oak Ridge Institute for Science and Education (ORISE) through an interagency agreement between the US Department of Energy and the Office of the Director of National Intelligence (ODNI).
\textbf{Author contributions:} J.R., M.F., A.I.A., C.F., M.C.C., S.K., and N.P.d.L.\ developed the theoretical framework for covariance magnetometry. J.R., Z.Y., and L.F.\ carried out covariance magnetometry experiments. J.R., C.F., M.C.C., S.K., and N.P.d.L.\ conceived the sensing technique, designed experiments, analyzed the data, and wrote the manuscript.
\textbf{Competing interests:} The authors declare no competing interests.
\end{acknowledgments}

\end{bibunit}


\pagebreak

\widetext
\begin{center}
\textbf{\large Supplementary Information}
\end{center}
\setcounter{secnumdepth}{3}
\setcounter{equation}{0}
\setcounter{figure}{0}
\setcounter{table}{0}
\setcounter{page}{1}
\makeatletter
\renewcommand{\theequation}{S\arabic{equation}}
\renewcommand{\thefigure}{S\arabic{figure}}
\renewcommand{\bibnumfmt}[1]{[S#1]}
\renewcommand{\citenumfont}[1]{S#1}
\begin{bibunit}[apsrev4-2]

\section{Methods}

The diamond sample was implanted with a nitrogen ion energy of 3 keV, resulting in shallow NV centers roughly $10\,$nm from the surface. NV center measurements are performed in a home-built dual-path confocal microscope setup. The green illumination on both paths is provided by a 532 nm optically pumped solid-state laser (Coherent Sapphire LP 532-300), split with a 50:50 beamsplitter (Thorlabs CCM5-BS016). Each path is then optically modulated by a dedicated acousto-optic modulator (AOM) (Isomet 1205C-1). The readout light around 590 nm is provided by different lasers for each path. The path 1 readout is provided by an NKT SuperK laser (repetition rate 78 MHz, pulse width 5 ps) with two bandpass filters with transmission wavelength around 590 nm (Thorlabs FB590-10 and Semrock FF01-589/18-25). The path 2 readout is provided by a 594 nm helium-neon laser (REO 39582). Both paths are optically modulated with dedicated AOMs (Isomet 1205C-1). The ionization light is provided by two 638 nm lasers (Hubner Cobolt 06-MLD) for each path, internally modulated. 

For each optical path, the three excitation wavelengths are combined by a 3-channel fiber RGB combiner (Thorlabs RGB26HF), and each excitation path is scanned by dedicated X-Y galvo mirrors (Thorlabs GVS012). The two optical paths are combined with a 2 inch beamsplitter cube (Thorlabs BS031). Each path is equipped with a 650 nm longpass dichroic mirror (Thorlabs DMLP650) to separate the excitation and collection pathways, and the photoluminescence (PL) for each path is measured by a dedicated fiber-coupled avalanche photodiode (Excelitas SPCM-AQRH-16-FC). A Nikon Plan Fluor 100x, NA = 1.30, oil immersion objective is used for focusing the excitation lasers and collecting the PL. The laser powers used (as measured before the objective) were approximately 3 to 7 $\mu$W for orange readout, 100 to 130 $\mu$W for green initialization, and 10 to 30 mW for the red ionization (this ionization power was extrapolated from lower-power measurements, and assumes perfect laser linearity). In practice, we found that the use of a green shelving pulse before ionization was unnecessary to achieve low readout noise, so a shelving pulse was not used.

Microwave pulses are generated using a Rohde and Schwarz signal generator (SMATE200A) and amplified with a high power amplifier (Mini-Circuits ZHL-16W-43S+) before being sent to a homemade microwave stripline. Low frequency test signals are generated with an arbitrary waveform generator (Keysight 33622A) and amplified with a high power amplifier (Mini-Circuits LZY-22+). For the data shown in \cref{fig:ACcorrelations}, we apply a random-phase AC signal at $f_0=2\,$MHz, phase-randomized with 1 MHz bandwidth Gaussian noise, which we detect using an XY8 dynamical decoupling sequence repeated 4 times (32 total pulses). For the data shown in \cref{fig:spectraldecomposition}A, we apply a $f_0=1.75\,$MHz AC signal phase-randomized with 50 kHz bandwidth Gaussian noise, detected with an XY8 sequence repeated 5 times (40 total pulses). For \cref{fig:spectraldecomposition}B we apply spectrally flat Gaussian noise with 2 MHz bandwidth and repeat an XY8 sequence twice (16 total pulses). For the data shown in \cref{fig:timeoffset}, the XY8 sequence is repeated twice (16 total pulses), and we measure an externally applied AC field at frequency $f_0=3.125\,$MHz using pulses separated by $\tau=160\,$ns. The AC signal is either phase-coherent (\cref{fig:timeoffset}C) or phase-randomized with 1 MHz bandwidth white noise (\cref{fig:timeoffset}D).

The correlation data were obtained by performing typically 1 to 2 million individual experiments for each data point shown, then correlating the resulting individual photon counts between the two paths. To filter out spurious correlations from slow PL variations due to sample drift, we subtract the mean photon number calculated for each 1000 data points sequentially, effectively high-pass filtering the raw counts. While this can help reduce spurious correlations from any significant background drifts in principle, the resulting change was minor in our data. 

\section{Temporal correlations between subsequent measurements}

\begin{figure}[ht]
	\centering
	\includegraphics[width=4in]{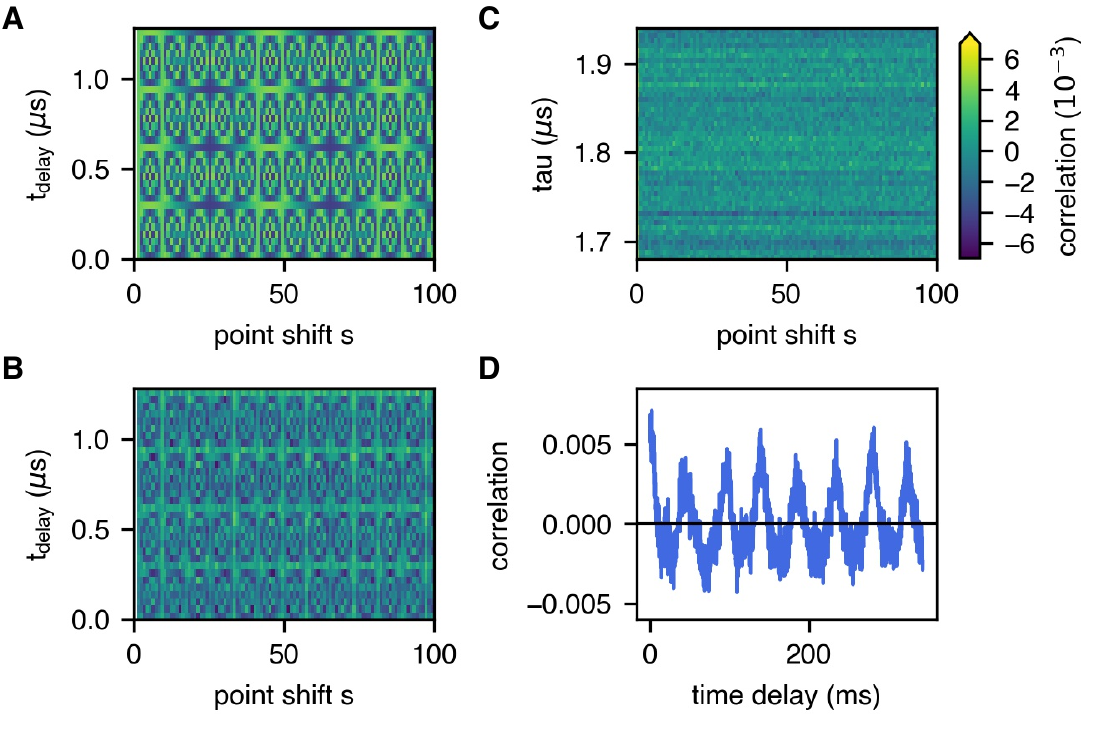}
	\caption{Temporal correlations between subsequent measurements. (A) Expected and (B) measured correlation in the signals from a 3.125 MHz source detected by 2 NV centers using XY8 sequences separated by $t_\text{delay}$. Because the experiment duration is milliseconds, the pattern is strongly aliased. (C) Measured correlation between signals from a white noise source, which drops into the noise for $s>0$. (D) Spurious correlation induced by roughly $20\,$Hz mechanical vibrations of the optical table.}
	\label{fig:supp_fig03}
\end{figure}

Covariance magnetometry with multiple NV centers enables correlation sensing with high temporal resolution, but we are further able to access the temporal correlation function between subsequent measurements $r(s) = \text{Cov}\left[S_1(i)S_2(i+s)\right]/(\sigma_1\sigma_2)$, where $s$ defines a relative offset and where the covariance is taken over the index $i$. For coherent AC signals like the ones measured in \cref{fig:timeoffset}B, we expect to see a temporal structure which depends on the pulse sequence duration and the signal frequency if the signal is stable for long periods of time (\cref{fig:supp_fig03}A). Although we did not set out to synchronize subsequent experiments with a stable clock, we are still able to observe this temporal structure in our data (\cref{fig:supp_fig03}B). The only free parameter in \cref{fig:supp_fig03}A is an overall offset in the experiment duration, which we set to 60 ns -- this is possibly caused by clock instabilities during the long charge state readout, which lasts a few milliseconds. For white noise signals like the ones measured in \cref{fig:spectraldecomposition}C, we instead expect the shot-to-shot correlations to be zero for $s>0$, which we also observe (\cref{fig:supp_fig03}C).

These long term time dynamics can also be useful for diagnosing experimental noise sources, which can cause significant problems in detecting true shot-to-shot correlations of the NV centers' spin states. As an example, \cref{fig:supp_fig03}D shows correlations which mimic a real spin signal but are in fact caused by mechanical vibrations due to a lateral contact point in one of the optical table legs. This vibration creates a global fluctuation in fluorescence collection on both optical paths and thus appears in the correlated signal. Removing such systematic noise sources is crucial for mitigating spurious correlations.

\section{Detectable correlations}

Consider two NV centers, which are not directly interacting with each other but which experience a shared classical magnetic field in their vicinity, as illustrated in \cref{fig:overview}A. This field is referred to as the correlated field. Each NV also sees proximal magnetic fields (for instance from fluctuating local nuclear spins), which are unique to each NV center. This will be called the uncorrelated or local field.

We isolate an effective spin-1/2 system for each NV by considering only the $m_s=0$ and $m_s=+1$ (or $-1$) sublevels, which we refer to as states 0 and 1 respectively. We assume each NV is initialized in the transverse plane, then in the course of some detection protocol each NV acquires some net transverse phase $\phi$ due to interactions with the magnetic field. At the end of the sensing protocol a final $\pi/2$ pulse maps the azimuthal angle $\phi$ to a polar angle $\theta=\phi + \pi/2$. Due to quantum projection the conditional probability for each NV center to be found in a given quantum state upon measurement is
\begin{align}
    P(m_s=0|\theta) = \cos^2(\theta/2) \\
    P(m_s=1|\theta) = \sin^2(\theta/2)
\end{align}
The measured signals $\sx$ and $\sy$ will depend on the measurement type; for single shot thresholded (th) measurement they will be the inferred spin states $0$ or $1$, while for photon counting (pc) they will be the counted photon numbers $k$, resulting in the measurement probabilities conditioned on the spin state:
\begin{align}
    &P^\text{pc}(s\seq k|m_s) = \Pois(k,\alpha_{m_s}) \\
    &P^\text{th}(s\seq m_s|m_s) = F
\end{align}
where $\Pois(k,\alpha_{m_s})$ is a Poisson process with mean $\alpha_{m_s}$ determined by the spin state, and $0.5 \leq F \leq 1$ is the readout fidelity. Below we will start by treating the thresholded readout more generally, allowing the error probabilities $P(s=0|m_s=1)$ and $P(s=1|m_s=0)$ to differ.

We repeat the experiment many times such that for each NV center we have a list of phases $\phi_i$ and corresponding measurements $s_i$
\begin{align}
    \Phix &= \{\phi_{1,i}\} \hspace{8mm} \Sx=\{\sxi\} \\
    \Phiy &= \{\phi_{2,i}\} \hspace{8mm} \Sy=\{\syi\}
\end{align}
where $i=1...N$ indexes the $N$ total experiments. The values of the $s_i$ will depend on whether we use photon counting or thresholded measurement as described above.

We are interested in what we can learn from the correlation between the two data sets $\Sx$ and $\Sy$ given certain assumptions about the distributions $\Phix$ and $\Phiy$. We will focus on the measured Pearson correlation $\rxy$, which will differ from the true statistical correlation due to quantum projection, finite sampling size, and readout error. In the following, we will use a Bayesian statistical model to derive the measured correlation between two such data sets, and find the sensitivity of such a measurement in different contexts.

\subsection{Ideal measured correlation}

We will not make any assumptions about the precise distributions from the correlated and local signals, except to assume that they are evenly distributed. We further assume that the phases acquired by the two NVs are
\begin{align}
    \phix = \phiGx + \phiLx \\
    \phiy = \phiGy + \phiLy
\end{align}
where $\phiGx,\phiGy$ are the common phases acquired due to the shared correlated field, and $\phiLx,\phiLy$ are the phases acquired due to the local field. We assume that $\phiGx\propto\phiGy$, and assume that $\phiLx$ and $\phiLy$ are independent. Such a decomposition into correlated and uncorrelated components is always possible for two lists, and here we take the correlated component $\phiGx,\phiGy$ to be caused by the global (shared) signal and the uncorrelated component $\phiLx,\phiLy$ to be caused by the local (unshared) signal.

The quantity we seek to derive is the Pearson correlation, defined by:
\begin{align}
    \rxy=\frac{\text{Cov}(\Sx,\Sy)}{\sigmaSx\sigmaSy}=\frac{\braket{\Sx \Sy}-\braket{\Sx}\braket{\Sy}}{\sigmaSx\sigmaSy}.
\end{align}
We start by assuming perfect readout so that the signals are the NV spin states $\{S_1,S_2\}=\{m_{s_1},m_{s_2}\}$. We let $p_\phi(\phix,\phiy)$ denote the joint probability density to acquire phase $\{\phix,\phiy\}$ with NV 1 and 2 respectively, and $p_s(s_1,s_2)=\int p(s_1|\phix)p(s_2|\phiy) p_\phi(\phix,\phiy) d\phix d\phiy$ denote the probability to detect signal $\{s_1,s_2\}$, where $p(s_i|\phi_i)$ is the probability to detect signal $s_i$ given accumulated phase $\phi_i$. Since we acquire phase in the transverse plane and read out after a final $\pi/2$ pulse (see \cref{fig:overview}) we have $\theta = \phi+\pi/2$, and
\begin{align}
    \braket{m_{s_1} m_{s_2}} &= 
    1\cdot 1 \cdot p_s(1,1) + 1\cdot 0 \cdot p_s(1,0) + 0\cdot 1 \cdot p_s(0,1) + 0\cdot 0 \cdot p_s(0,0) \nonumber \\
    &= p_s(1,1) \\
    &= \int_{\phix,\phiy}
    \sin^2{\left(\frac{\phix}{2}+\frac{\pi}{4}\right)} \sin^2{\left(\frac{\phiy}{2}+\frac{\pi}{4}\right)} p_\phi(\phix, \phiy) d\phix d\phiy
\end{align}
Since we assume $\phiG, \phiLx, \phiLy$ are drawn from independent distributions, we may rewrite the phase probabilities in terms of separate statistical draws, and we have
\begin{align}
    &\braket{m_{s_1} m_{s_2}} = \nonumber \\ &\int
    \sin^2{\left(\frac{1}{2}[\phiGx + \phiLx]+\frac{\pi}{4}\right)} \sin^2{\left(\frac{1}{2}[\phiGy + \phiLy]+\frac{\pi}{4}\right)} \, p(\phiG) p(\phiLx) p(\phiLy) \, d\phiG d\phiLx d\phiLy \\
    &= \frac{1}{4}(1 + \braket{\sin(\phiGx)\sin(\phiGy)}
    \braket{\cos(\phiLx)}\braket{\cos(\phiLy)}) \label{eq:meanXY}
\end{align}
where we have used the fact that $\braket{\sin(\phi)}=0$ for an even distribution. Then we have for the correlation (using the Bernoulli statistics $\braket{S}=\braket{S^2}=1/2$ and $\sigma_S^2 = \braket{S^2} - \braket{S}^2 = 1/4$)
\begin{align}
    \rideal = \braket{\sin(\phiGx)\sin(\phiGy)}
    \braket{\cos(\phiLx)}\braket{\cos(\phiLy)}
\end{align}

Noticing that $\braket{\cos(\phiL)} = \braket{\e^{i\phiL}} = \e^{-\tilde{\chi}(t)}$ is the decoherence function for variance detection \cite{Degen2017}, we find the correlation 
\begin{align}
    \rideal = \e^{-[\tilde{\chi}_1(t)+\tilde{\chi}_2(t)]} \braket{\sin[\phiGx(t)]\sin[\phiGy(t)]} \label{eq:ridealSI}
\end{align}
Note that \Cref{eq:ridealSI} is similar to the expression for temporal correlation spectroscopy using a single NV center \cite{Laraoui2013}, in which case there are two subsequent phase acquisition times for the single NV center instead of independent phase acquisition times for two separate NV centers. 

As an example, for identical correlated phases $\phiGx=\phiGy=\phiG$ this is
\begin{align}
    \rideal = \frac{1}{2}\e^{-[\tilde{\chi}_1(t)+\tilde{\chi}_2(t)]} \left[1-\braket{\cos(2\phiG(t))}\right] \label{eq:ridealSIequalphase}
\end{align}
which for Gaussian-distributed correlated phases is
\begin{align}
    \rideal = \frac{1}{2}\e^{-[\tilde{\chi}_1(t)+\tilde{\chi}_2(t)]} \left(1-\e^{-2\sigmaG^2} \right)
\end{align}
where $\sigmaG^2$ is the variance of the correlated phase distribution. The assumption of Gaussian phases is violated for e.g.\ random-phase AC signals detected by a CP-type pulse sequence with pulse spacing $\tau$, in which case the more general Bessel function forms will result \cite{Degen2017} if the correlated phases are identical:
\begin{align}
    \rideal = \frac{1}{2}\e^{-(\tilde{\chi}_1(t)+\tilde{\chi}_2(t))}\left[1-J_0\left(4\gamma B_0 \overline{W}t\right)\right] \label{eq:BesselRhoSupp},
\end{align}
where $\overline{W}=\text{sinc}(\pi f n \tau)\left[1-\sec(\pi f \tau)\right]$ and $n$ is the total number of applied pulses. Note the extra factor of 2 relative to the expression for decoherence measured using typical single-NV center variance detection \cite{Degen2017}. While decoherence is effectively accelerated because of contributions from both NV centers (since there are two factors of $\tilde{\chi}$ in \cref{eq:ridealSI}), the phase accumulation rate is also effectively doubled (since $\sin^2(\phi)\sim \cos(2\phi)$), such that there is no net penalty to the sensitivity regarding phase integration time.

\subsection{Measured correlation with readout noise}

\subsubsection{Photon counting: shot noise}

We are now interested in accounting explicitly for the number of photons $n$ that are counted from an NV center, depending on its state. The photon number is drawn from a Poisson distribution whose mean depends on the NV center spin state, with mean $\alpha_0$ for state $m_s=0$ and mean $\alpha_1$ for $m_s=1$.

The list of photon counts for NV 1 is $\Sx$ and for NV 2 is $\Sy$, with individual photon counts $\nx$ and $\ny$. As before, we must calculate $\braket{\Sx \Sy} = \sum  \nx \ny P(\nx,\ny)$:
\begin{align}
    \braket{\Sx \Sy} = \sum_{\nx,\ny} \nx \ny \big[ 
    &P(\nx|m_s\seq0)P(\ny|m_s\seq0) \int_{\phi_1,\phi_2}\cos^2{\left(\frac{\phi_1}{2}+\frac{\pi}{4}\right)} \cos^2{\left(\frac{\phi_2}{2}+\frac{\pi}{4}\right)} p(\phi_1, \phi_2) d\phi_1 d\phi_2 + \nonumber \\
    &P(\nx|m_s\seq0)P(\ny|m_s\seq1) \int_{\phi_1,\phi_2}\cos^2{\left(\frac{\phi_1}{2}+\frac{\pi}{4}\right)} \sin^2{\left(\frac{\phi_2}{2}+\frac{\pi}{4}\right)} p(\phi_1, \phi_2) d\phi_1 d\phi_2 + \nonumber \\
    &P(\nx|m_s\seq1)P(\ny|m_s\seq0) \int_{\phi_1,\phi_2}\sin^2{\left(\frac{\phi_1}{2}+\frac{\pi}{4}\right)} \cos^2{\left(\frac{\phi_2}{2}+\frac{\pi}{4}\right)} p(\phi_1, \phi_2) d\phi_1 d\phi_2 + \nonumber \\
    &P(\nx|m_s\seq1)P(\ny|m_s\seq1) \int_{\phi_1,\phi_2}\sin^2{\left(\frac{\phi_1}{2}+\frac{\pi}{4}\right)} \sin^2{\left(\frac{\phi_2}{2}+\frac{\pi}{4}\right)} p(\phi_1, \phi_2) d\phi_1 d\phi_2 \big],
\end{align}
or recognizing the angular integral from above, and using by symmetry
\begin{align}
p(m_s\seq0)=p(m_s\seq1)= \int \sin^2{\left(\frac{\phi_i}{2}+\frac{\pi}{4}\right)} p(\phi_i) d\phi_i = \int \cos^2{\left(\frac{\phi_i}{2}+\frac{\pi}{4}\right)} p(\phi_i) d\phi_i = \frac{1}{2},
\end{align}
we have
\begin{align}
    \braket{\Sx \Sy} &= \sum_{\nx,\ny} \nx \ny \times \nonumber \\
    \big[ &P(\nx|m_s\seq0)P(\ny|m_s\seq0) \braket{\Xms \Yms} + \nonumber \\
    &P(\nx|m_s\seq0)P(\ny|m_s\seq1) \left(\tfrac{1}{2}- \braket{\Xms \Yms}\right) + \nonumber \\
   &P(\nx|m_s\seq1)P(\ny|m_s\seq0) \left(\tfrac{1}{2}- \braket{\Xms \Yms}\right) + \nonumber \\
    &P(\nx|m_s\seq1)P(\ny|m_s\seq1)  \braket{\Xms \Yms} \big]
\end{align}
where $\braket{\Xms \Yms}$ is defined as in \cref{eq:meanXY}. Because the draws for $\nx$ and $\ny$ are independent at this stage (i.e.\ $\braket{\nx\ny}=\braket{\nx}\braket{\ny}$ when drawn from already-given Poisson distributions), and denoting a Poisson distribution with mean $\alpha$ as $\text{Pois}_\alpha(x)$, we can use e.g.\ $\sum_{\nx} \nx P(\nx|m_s\seq 0) = \sum_{\nx} \nx \text{Pois}_{\alpha_0}(\nx) = \alpha_0$ to find
\begin{align}
    \braket{\Sx \Sy} = \braket{\Xms \Yms}(\alpha_0-\alpha_1)^2 + \alpha_0\alpha_1.
\end{align}
Lastly, we note that for a joint Poisson distribution we have mean and variance:
\begin{align}
    \braket{S} &= \frac{1}{2}(\alpha_0 + \alpha_1) \\
    \braket{S^2}-\braket{S}^2 &= \frac{1}{4}(\alpha_0-\alpha_1)^2 + \frac{1}{2}(\alpha_0 + \alpha_1). 
\end{align}
Combining these elements the detected correlation for photon counting $\rPC$ becomes
\begin{align}
    \rPC &= \frac{1}
    {1+2(\alpha_0+\alpha_1)/(\alpha_0-\alpha_1)^2} \rideal \nonumber \\
    &= \frac{1}{\sigma_R^2} \rideal,
\end{align}
where $\sigma_R=\sqrt{1+2(\alpha_0+\alpha_1)/(\alpha_0-\alpha_1)^2}$ is the readout noise \cite{Taylor2008,Hopper2018}. Notice that the measured correlation depends quadratically on the readout noise, rather than linearly. 

If we assume that the two NV centers have different readout noise $\sigma_{R_1}$ and $\sigma_{R_2}$, a slightly longer but straightforward calculation yields the more general result:
\begin{align}
    \rPC &= \frac{1}{\sigma_{R_1}\sigma_{R_2}} \rideal, \label{eq:rpcSupp}
\end{align}

\subsubsection{Single shot readout: thresholding}

For a thresholded measurement with $P_k(i,j)$ the probability to assign spin state $i$ given spin state $j$ on NV center $k$, we can perform a similar calculation to the one above to find:
\begin{align}
    \braket{\Sx \Sy} =& P(\Sx\seq 1,\Sy\seq 1)  \nonumber \\
    =& P_1(1|0)P_2(1|0) \braket{\Xms\Yms} + \nonumber \\
    =& P_1(1|0)P_2(1|1) \left(\tfrac{1}{2}-\braket{\Xms\Yms}\right) + \nonumber \\
    =& P_1(1|1)P_2(1|0) \left(\tfrac{1}{2}-\braket{\Xms\Yms}\right) + \nonumber \\
    =& P_1(1|1)P_2(1|1) \braket{\Xms\Yms}.
\end{align}
Then, since $\braket{S_i^2}=\braket{S_i}=\tfrac{1}{2}\left[P_i(1|0)+P_i(1|1)\right]$, we find the detected correlation for thresholding $\rSS$
\begin{align}
    \rSS=\frac{1}{\sigma_{R_1}^\text{th}\sigma_{R_2}^\text{th}}\rideal
\end{align}
where the readout noise for thresholding is \cite{Hopper2018}
\begin{align}
    \sigma_{R_i}^\text{th}=\sqrt{1+2\frac{P_i(1|0)\left[1-P_i(1|0)\right]+P_i(1|1)\left[1-P_i(1|1)\right]}{\left[P_i(1|0)-P_i(1|1)\right]^2}}.
\end{align}
In the simplified case that the errors are symmetric with $P(1|0)=1-P(1|1)$ we have $\sigma_{R_i}^\text{th}=1/(2F-1)$ where $F=1-\tfrac{1}{2}\left[P(1|0)+\left[1-P(1|1)\right]\right]\rightarrow 1-P(1|0)$ is the fidelity \cite{Hopper2018}. The detectable correlation then becomes:
\begin{align}
    \rSS=(2F_1-1)(2F_2-1)\rideal,
\end{align}
where the two NV centers may have different readout fidelities $F_1$ and $F_2$ respectively. When fidelity is minimized ($F=0.5$) the measured correlation is 0, and when fidelity is maximized ($F=1$) we recover the idealized correlation $\rideal$ in \cref{eq:ridealSI}. Again, note that the measurable correlation depends quadratically on the readout fidelity rather than linearly.

\section{Expected correlations}

\begin{figure*}[ht]
	\centering
	\includegraphics[width=\textwidth]{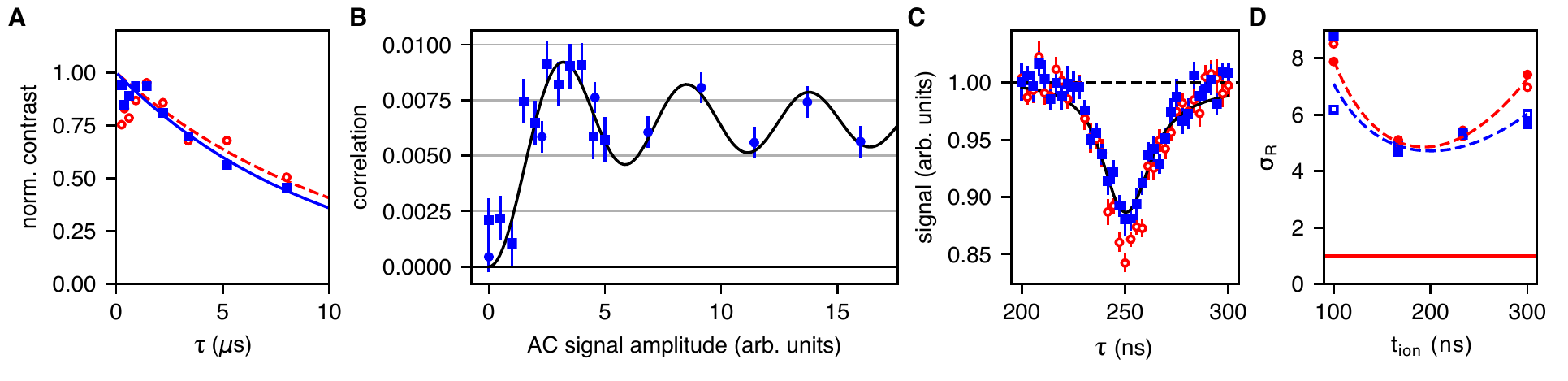}
	\caption{Measuring parameters in estimating expected correlation for \cref{fig:ACcorrelations}. (A) Dynamical decoupling measurements provide decoherence rates with $\text{exp}(-t/T_2)\approx 0.94$ and $0.71$ for the two NV centers near $\tau=250\,$ns. (B) Correlation measurements versus amplitude provide an estimate of the AC signal magnetic field strength $B_0\approx 0.13\,$G. (C) XY8 measurements provide an estimate of the FWHM of the 2 MHz signal with 1 MHz phase noise. In \cref{eq:BesselRhoSupp}, we integrate $f$ over the frequency range of our line-broadened signal to derive the theory curves shown in \cref{fig:ACcorrelations}B. (D) Readout noise $\sigma_R$ versus ionization time measured before (open markers) and after (filled markers) the data acquisition in \cref{fig:ACcorrelations}. Lines are guides to the eye. $t_\text{ion}=200\,$ns was used to acquire the data shown in \cref{fig:ACcorrelations}.}
	\label{fig:supp_fig02}
\end{figure*}

We estimate the expected detectable correlations in our experiment by measuring each of the key parameters in \cref{eq:BesselRhoSupp,eq:rpcSupp}: decoherence, magnetic field strength, frequency range, and readout noise. This characterization is show in \cref{fig:supp_fig02} for each of these variables in turn. Numerically calculating the expected correlation using these variables we find that our measured correlation is approximately 70\,\% of the expected value, likely limited by charge state initialization and SCC ionization efficiency.

Decoherence from local sources reduces the measurable correlation, as shown in \cref{fig:spectraldecomposition}A,C. In both cases the local noise is due to the hyperfine interaction of the NV center with its intrinsic nuclear spin. The filter function frequencies where this interaction is detected are at  \cite{Abe2018} $f_k = (2 \gamma_N B_0 + 3.05 \,\text{MHz} )/(2 k)$, where $\gamma_N =-4.3\,$MHz/T is the $^{15}$N nuclear gyromagnetic ratio, $B_0\approx 31\,$mT is the strength of the external magnetic field, and $k$ is the filter function frequency harmonic. In \cref{fig:spectraldecomposition}A,C in the main text, the detected harmonics are $k=1$ and $k=5$ respectively.
For the data shown in \cref{fig:timeoffset}, note that the $0\rightarrow+1$ detuning is only about $60\,$MHz due to hybridization for the misaligned NV center \cite{Epstein2005}.

\section{Spectral decomposition and sensitivity}

\subsection{Spectral decomposition}

We assume the NV centers accumulate small phase angles (or experience a Gaussian noise source) in the presence of a Carr-Purcell (CP) \cite{Carr1954} type AC sensing sequence with large pulse numbers, and pulse spacing $\tau$. Approximating the pulse sequence filter function as a delta function centered at frequency $\omega=\pi/\tau$, the coherence decay $C(t)$ is generally described by
\begin{align}
     C(t) &= \e^{-\chi(t)} \nonumber \\
     \chi(t) &= \frac{1}{2}\braket{\phi^2} = 
     \frac{1}{\pi}\int_0^{\infty} d\omega \frac{F(\omega)}{\omega^2}S(\omega)
     \approx \frac{t}{\pi} S(\omega), \label{eq:spectralbasics}
\end{align}
where $F(\omega)$ is the pulse sequence filter function, $\chi(t)$ is the decoherence from \emph{all} noise sources (local and global), and $S(\omega)$ is the spectral density of the magnetic field \cite{Degen2017,Szankowski2017,Romach2015}:
 \begin{align}
     S(\omega) & = \int_{-\infty}^{\infty}\e^{-i\omega t}\gamma_e^2 G(t)dt   \\
     G(t) &= \braket{B(t'+t)B(t')}.
\end{align}

To perform spectral decomposition we assume that the two NV centers experience identical global fields with noise spectral densities $S_C(\omega) = S_{C_1}(\omega) = S_{C_2}(\omega)$, and that the noise spectrum may be decomposed into correlated and uncorrelated (local) contributions $S(\omega) = S_C(\omega)+S_{L_{1,2}}(\omega)$. We further assume that the accumulated correlated phases are Gaussian-distributed or small such that $\phi\ll\pi$. Then we have $\braket{\sin(\phiGx)\sin(\phiGy)}=\e^{-2\chi_C}\sinh(2\chi_C)$, where $\chi_C$ is the decoherence induced by the correlated noise source, and the correlation becomes
\begin{align}
     r=\frac{1}{\sigma_R^2}C_1(t)C_2(t)
     \sinh\left(\frac{2 t}{\pi}S_C\left(\frac{\pi}{\tau}\right)\right), \label{eq:rGaussian}
\end{align}
where $C(t)=\e^{-(\tilde{\chi}+\chi_C)t}$ is the total decoherence from all sources. Inverting this equation we find the correlated spectral density 
\begin{align}
    S_C(\omega) = \frac{\pi}{2t}
    \sinh^{-1}\left(\frac{\sigma_R^2 r}{C_1(t) C_2(t)}\right). \label{eq:reconstructedSpectrumSupp}
\end{align}
which is \cref{eq:reconstructedSpectrum} in the main text.

\subsection{Sensitivity}

To derive the sensitivity of a covariance magnetometry measurement, we start from \cref{eq:rpc} in the main text, which accounts for the signal, readout noise, and decoherence. We now account for the statistical noise $\varsigma_r$ which is a measure of the uncertainty in the Pearson correlation due to the finite number of sampled points $N$ \cite{Fisher1925}:
\begin{align}
    \varsigma_r \approx \tanh\left(\frac{1}{\sqrt{N-3}}\right)\approx \frac{1}{\sqrt{N}},
\end{align}
where the approximation holds for $N\gg1$. Then the SNR is approximately
\begin{align}
    \text{SNR} = \frac{r}{\varsigma_r} \approx \frac{\e^{-2\tilde{\chi}(t)}\sqrt{N}}{\sigma_R^2} \braket{\sin[\phiGx(t)]\sin[\phiGy(t)]}. \label{eq:SNR}
\end{align}
For simplicity, we have assumed that the NV centers have the same readout noise $\sigma_R$ and the same decoherence function from local noise sources $\tilde{\chi}(t)$.

To determine the sensitivity we must consider the time dependence of each term in \cref{eq:SNR}. These include the time it takes to run each of the $N$ experiments, the phase accumulation time, and potentially the time dependence of the readout noise $\sigma_R$ (which for SCC improves for longer readout times). We assume the detected correlations are from a shared magnetic field source with spectral density $S_C(\omega)$ (\cref{eq:rGaussian}), where we again assume that the two NV centers see the same shared field (rather than e.g.\ one NV center being further and experiencing an attenuated version of the shared field), so that $S_{C_1}(\omega)=S_{C_2}(\omega)=S_{C}(\omega)$. Then
\begin{align}
    S_{C,\text{min}}
    &=-\frac{\pi}{4t} \ln \left[1-\frac{2\sigma_R^2 \e^{\chi_{L1}(t)+\chi_{L2}(t)}}{\sqrt{N}}\right] \nonumber \\
    &\approx \frac{\pi}{2}\sigma_R^2\e^{2t/T_2} \sqrt{\frac{t+t_\text{R}}{t^2 T}},
\end{align}
where $t$ is the phase integration time, $t_\text{R}$ is the readout time, and $T\approx (t+t_\text{R})N$ is the total experiment time ignoring initialization. Assuming the noise has flat spectral density around the detection frequency $S_{C}(\omega)=\gamma_e^2\sigma_B^2/\text{Hz}$ we find the minimum detectable noise amplitude
\begin{align}
    \sigma_{B,\text{min}}^2 &= \frac{-\pi\cdot\text{Hz}}{4\gamma_e^2t} \ln \left( 1-\frac{2\sigma_R^2 \e^{2t/T_2}}{\sqrt{T/(t+t_\text{R})}} \right), \label{eq:sensitivitygeneralSupp}
\end{align}
which is \cref{eq:sensitivitygeneral1} in the main text and is illustrated in \cref{fig:ACcorrelations}C.

\section{Higher order joint cumulants}

We have focused on 2-body (Pearson) correlations but here we extend this to higher orders. Consider the $N$th-order joint cumulant defined by 
\begin{align}
    \kappa_N=\kappa(m_1,m_2,...,m_N)=
    \sum_\pi (|\pi|-1)! (-1)^{|\pi|-1}
    \Pi_{B \in \pi} \braket{\Pi_{i\in B} m_i} \label{eq:jointcumulants}
\end{align}
where $\pi$ are the different partitions (ways of grouping the individual $m_i$), $|\pi|$ is the number of parts in a partition, and $B$ is the blocks in the partitions. For example, for $N=3$ we have 
\begin{align}
    \kappa_3 = \braket{m_1,m_2,m_3} - \braket{m_1,m_2}\braket{m_3} -
    \braket{m_1,m_3}\braket{m_2} - 
    \braket{m_2,m_3}\braket{m_1} + 
    2 \braket{m_1,m_2,m_3},
\end{align}
where e.g.\ the partition $\braket{m_1,m_2}\braket{m_3}$ has two blocks ($|\pi|=2$), which are $\braket{m_1,m_2}$ and $\braket{m_3}$.
Here we calculate this joint cumulant for $N$ NV centers where we assume each NV center experiences the same magnetic field for simplicity.

Suppose we arrange our starting NV center orientations from measurement to measurement in such a way that across many measurements all NV measurement expectation values are independent; for instance with four NV centers we have $\braket{m_1m_2m_3}=\braket{m_1}\braket{m_2}\braket{m_3}$, etc., where for a Bernoulli distribution with NV states $m_i=0$ or $1$ we have $\braket{m_i}=1/2$. Then in \cref{eq:jointcumulants} we must calculate $\braket{m_1m_2...m_N}$ as well as a series of terms which will only contain products of individual means $\braket{m_i}$:
\begin{align}
    \kappa_N = \braket{m_1m_2...m_N} + \left(\sum_i x_i\right) \braket{m}^N \label{eq:cumulantdeduce}
\end{align}
where $x_i$ are coefficients to the series of mean-product partition terms in \cref{eq:jointcumulants}. 

However, the latter term may be quickly deduced by noticing that for any cumulant with independent entries we must have
\begin{align}
    \kappa_N^\text{indep}=0&=\braket{m_1m_2...m_N} + \left(\sum_i x_i\right)\braket{m_1}\braket{m_2}...\braket{m_N} \nonumber \\
    &=\left(1+\sum_i x_i\right)\braket{m_1}\braket{m_2}...\braket{m_N}
\end{align}
so that $\sum_ix_i=-1$ and the second term in \cref{eq:cumulantdeduce} must be $-\braket{m}^N=-1/2^N$. 

For the first term we have
\begin{align}
    \braket{m_1m_2...m_N} = 1\cdot p(1,1,...,1) &= \frac{1}{2^N}\int (1+\sin(\phi))^N p(\phi) d\phi \nonumber \\
    &= \frac{1}{2^N} \int 1+\sin^N(\phi) p(\phi) d\phi
\end{align}
where the last equality holds because we have already assumed the phase distributions are independent for any number of NVs $m<N$. Then for the $N$th-order cumulant we have
\begin{align}
    \kappa_N &= \frac{1}{2^N}\int 1 + \sin^N(\phi) p(\phi) d\phi - \frac{1}{2^N} \nonumber \\
    &= \frac{1}{2^N}\braket{\sin^N(\phi)}
\end{align}

Lastly, by analogy with the usual expression for the Pearson correlation, we define a normalized $N$th-order joint cumulant $\tilde{\kappa}_N$ by 
\begin{align}
  \tilde{\kappa}_N = \frac{\kappa_N}{\Pi_i \sigma_i}  
\end{align}
where $\sigma_i$ are the standard deviations of the individual distributions; for Bernoulli distributions these are $\sigma_i = 1/2$ yielding
\begin{align}
    \tilde{\kappa}_N = 2^N \kappa_N = \braket{\sin^N(\phi)}.
\end{align}
The Fourier term in this expression whose rate is $N\phi$ is suppressed by a factor $1/2^N$, such that the net sensitivity relative to single-NV variance sensing is $\sqrt{N}/2^{N-1}$ for this term. Thus the boosted phase accumulation rate does not translate to an overall sensitivity enhancement for large $N$ relative to single-NV sensing of the same signal.

\putbib[Correlated_sensing]

\end{bibunit}

\end{document}